# PERTURBATION THEORY IN A MICROCANONICAL ENSEMBLE

Ritapriya Pradhan    and    Jayanta K. Bhattacharjee

School of Physical Sciences

Indian Association for the Cultivation of Science

Jadavpur        ,        Kolkata 700032

INDIA

## Abstract

The microcanonical ensemble is a natural starting point of statistical mechanics. However, when it comes to perturbation theory in statistical mechanics, traditionally only the canonical and grand canonical ensembles have been used. In this article we show how the microcanonical ensemble can be directly used to carry out perturbation theory for both non-interacting and interacting systems. We obtain the first non-trivial order answers for the specific heat of anharmonic oscillators and for the virial expansion in real gases. They are in exact agreement with the results obtained from the canonical ensemble. In addition, we show how crossover functions for the specific heat of anharmonic oscillators can be constructed using a microcanonical ensemble and also how the subsequent terms of the virial expansion can be obtained. However, we find that if we consider quantum free particles in a one-dimensional box of extension L, then the two ensembles give strikingly different answers for the first correction to the specific heat in the high temperature limit.

## I.    INTRODUCTION

The basic purpose of statistical mechanics [1-2] is to connect the dynamics of a very large number $N$ (typically of the order of Avogadro number) of particles with their thermodynamic properties like pressure, specific heat etc. Describing the dynamics of such a large number of particles necessitates a probabilistic description even when the interactions are purely elastic collisions. If the dynamics occurs in a D-dimensional space ( $D=1$ if restricted to a line, $D=2$ if confined in a plane, $D=3$ if confined in a volume and so on ), then the dynamical state of each particle is specified by $D$ components of the position vector and $D$ components of the momentum vector. Hence to specify the dynamical state of each particle one requires $2D$ numbers. For $N$ particles, one requires $2DN$ numbers to specify a state. The dynamical state of a system of $N$ particles is, consequently, represented by a point in a $2ND$ dimensional space known as phase space. The dynamics is portrayed by a curve in phase space – the succession of states as the system evolves according to Newton's laws.

To avoid extra mathematical complexities, we will confine ourselves to $D=1$ in Sections II and III and talk about the evolution of the system in a $2N$ dimensional phase space. The trajectory of the system evolves keeping the total energy conserved. Our system is a collection of non-interacting particles in an external potential $\phi(x)$. The energy of the $i-th$ particle is $E_i$ and is given by

$$E_i = \frac{p_i^2}{2m} + \phi(x_i) \tag{1.1}$$

and the total energy is $E = \sum_{i=1}^{i=N} E_i$. The trajectory of the system is confined on the surface of the hypersphere defined by the equation $E = \sum_{i=1}^{N}\left[\frac{p_i^2}{2m} + \phi(x_i)\right]$. In the presence of an interaction, however weak, the trajectory spreads out evenly over the surface of the hypersphere and the total number of microscopic states available to the system can be defined as the volume of the accessible region of the phase space divided out by the infinitesimal volume element $h^N$ where $h$ is Planck's constant. The division discretizes the continuum to a large number. The number of accessible states of a system of $N$ non-interacting particles confined on a one-dimensional line of extension $L$ and having total energy $E$ is

$$\Gamma(N,E,L) = \int \frac{dx_1 dp_1}{h} \frac{dx_2 dp_2}{h} \dots \frac{dx_N dp_N}{h} \delta\left( E - \sum_{i=1}^{N} [\frac{p_i^2}{2m} + \phi(x_i)] \right) \qquad (1.2)$$

The delta function in the above equation ensures that the orbit covers the surface of a volume in a $2N$ dimensional space. A very detailed discussion of the microcanonical ensemble is given in the first chapter of the statistical physics text of Landau and Lifshitz [1]. Our Eq. (1.2) is a small variation of Eq.(4.4) of Chapter 1 of this classic text. It should be noted that in certain cases an additional factor of $1/N!$ will be necessary to avoid Gibb's paradox and we will discuss this as we consider the different systems.

The fundamental postulate of statistical mechanics asserts that each of the above microstates is equally accessible. The connection with thermodynamics is achieved by Boltzmann's hypothesis

$$S = k \ln \Gamma(E,N,V) \qquad (1.3)$$

The temperature $T$ follows from

$$\frac{1}{T} = \frac{\partial S}{\partial E}\bigg|_{N,V} \qquad (1.4)$$

and all thermodynamic quantities can be subsequently obtained.

What we have just described is the procedure for using the microcanonical ensemble in statistical mechanics and could in principle cover a large part of the subject (an example of an important omission is the whole area of phase transitions). In practice this technique is difficult to implement. The microcanonical approach is used primarily to study the two-state system, the free particle and the simple harmonic oscillator. Usually, it not used for systems with an inherent scale (anharmonic oscillators) or interacting systems (e.g., the Van der Waals gas), although recently, it has been used in a couple of unconventional settings [3,4]. When practical applications are concerned, this ensemble is rarely used. Some notable exceptions are application to particles in a gravitational field [5], a particular formulation of lattice gauge theory [6] and a variant on Ref. [1] for the quantum case [7]. An interesting consequence of treating small systems in the microcanonical ensemble has been studied in Ref. [8]. For a discussion of Eq. (1.3) and the definition involving the probability of occupying a given microstate in the ensemble of choice, one should consult Ref. [9]. We will maintain the large $N$ limit throughout and hence not enter the interesting issues pointed out by Dunkel and Hilbert [10].

In this article we want to point out that although somewhat cumbersome, the use of the microcanonical ensemble can indeed be extended to the study of anharmonic oscillators (as an example of a non-interacting system with an inherent scale). Within perturbation theory, we calculate the specific heat of oscillators which are admixtures of different restoring forces. This is done in Secs II and III. What is particularly interesting is that within the microcanonical picture, we can exploit the approximation techniques of classical dynamics of anharmonic oscillators to actually calculate the crossover functions for the specific heat that takes us from the simple harmonic to the fully anharmonic limit. This is done in Sec IV.

An interesting issue in the microcanonical ensemble has been that of the negative specific heat for self-gravitating systems. The canonical ensemble result, $kT^2 C_V = \left\langle (E - \bar{E})^2 \right\rangle$, where $\bar{E}$ is the mean value of the energy $E$, shows that the specific heat $C_V$ can never be negative. However, canonical ensemble implies equilibrium with a bath at constant temperature $T$. The microcanonical ensemble does not have this constraint and it was shown by several authors [11-16] that in a microcanonical ensemble the specific heat of a self-gravitating system can indeed be negative. The apparent paradox was resolved by Thirring and Hertel [13] who set up a model system which has a negative specific heat for a range of energy values in the microcanonical ensemble and a phase transition in the corresponding canonical ensemble.

This history prompted us to look at the case of real gases with two-particle interactions (short range repulsion at very close range and weak attraction at large separations) within the microcanonical frame-work using the perturbation techniques developed in the previous sections. We find that the perturbation theory yields the same result for the equation of state as the canonical ensemble and is thus consistent with Van der Waals equation which is obtained if the expansion (virial expansion) of the pressure $P$ in powers of the number density $n$ is truncated at the second order. Interestingly enough, Van der Waals equation shows a phase transition from a one-phase to a two-phase region of negative susceptibility. This feature is obtained as a result of the truncation. In the usual canonical approach, it is known how to obtain the higher order terms of the virial expansion. That this is feasible in the micro-canonical expansion as well has been shown in Sec V. The inequivalence of the constant $E$ (microcanonical) and constant $T$ ( canonical ) ensembles seem to be related to the long range and short ranged interactions in a many body system. The actual result, as has been shown in Refs [17] and [18], is that the equivalence depends upon the concavity of the thermodynamic entropy. Our goal here is much more restricted. We want to point out that perturbation theory, a very standard tool for calculations and never really carried out in the microcanonical ensemble, can also be carried out under the constant energy constraint. However, we find an interesting difference in the results of the two ensembles when we consider a set of non-interacting quantum particles enclosed in a one-dimensional box of extension L. In the high temperature limit, both ensembles give the molar specific heat of R but when we consider the first correction, we find the corrections to be different both qualitatively and quantitatively in the two ensembles.

In Secs II and III, we will deal with the nonlinear oscillator whose single particle Hamiltonian is given by

$$H = \frac{p^2}{2m} + \frac{1}{2} m\omega^2 x^2 + \frac{\lambda}{4} x^4 \qquad (1.5)$$

At any temperature T, there is a dimensionless parameter in the system given by (this is the inherent scale in the system)

$$\mu = \frac{\sqrt{\lambda kT}}{m\omega^2} \qquad (1.6)$$

For $\mu \to 0$, we have the familiar simple harmonic oscillator with the molar specific heat $C_V = R$. On the other hand for $\mu \to \infty$, we have the quartic oscillator for which $C_V = 3R/4$. For usual values of the constants $m, \omega$ and $\lambda$, the specific heat at room temperature is very close to $R$. If we now raise the temperature, the specific heat will begin to fall and reach $3R/4$ for $\mu \gg 1$. The reverse trend will be seen if we start at very high temperatures and cool to the room temperature. The standard method of studying this temperature variation is to use the canonical ensemble. Calculating the first correction to the leading term in the two limits ($\mu \ll 1$ and $\mu \gg 1$ as recalled in Sec II), one finds for $\mu \ll 1$

$$C_V = R - O(\mu^2) \tag{1.7}$$

and for $\mu \gg 1$

$$C_V = \frac{3R}{4} + O\left(\mu^{-1}\right) \tag{1.8}$$

In Sec III, we will obtain the identical results from the microcanonical ensemble. In Sec IV we show how the microcanonical ensemble can be used to construct an explicit crossover function that takes us from Eq. (1.7) to Eq. (1.8) as the parameter $\mu$ is varied. In Sec V, we focus on the real gas with pairwise interaction and show that by carrying out a first order perturbation theory, we can arrive at the Van der Waals equation which is also retained at the second order. Beyond the second order it is possible to construct a virial expansion for the pressure but as in the canonical expansion, this virial expansion is not consistent with the Van der Waals equation of state. It should be pointed out that we will always work in the thermodynamic limit where $N \to \infty, E \to \infty$ but the ratio $E/N$ is always finite. In Sec VI, we discuss the one-dimensional quantum gas where, as already mentioned, the two ensembles give different answers. We conclude with a brief summary in Sec VII.

## II.    PERTURBATION THEORY IN THE CANONICAL ENSEMBLE

In this section we briefly recall how the specific heat of the anharmonic oscillator of Eq. (1.5) is calculated in the canonical ensemble. For a system of $N$ independent oscillators the partition function $Z_N$ is related to the single particle partition function $Z$ by the usual relation

$$Z_N = Z^N \qquad (2.1)$$

The partition function $Z$ is calculated in perturbation theory by writing (here we assume that the quadratic term dominates) as

$$Z = \int \frac{dp\,dx}{h} \exp\left[ -\left( \frac{p^2}{2m} + \frac{1}{2} m\omega^2 x^2 + \frac{\lambda}{4} x^4 \right) / kT \right]$$

$$= \left( \frac{2\pi mkT}{h^2} \right)^{1/2} \int_{-\infty}^{\infty} dx\, e^{-\left( m\omega^2 x^2 / 2kT \right)} \left[ 1 - \frac{\lambda x^4}{4kT} + O(\lambda^2) \right]$$

$$= \left( \frac{2\pi mkT}{h^2} \right)^{1/2} \left( \frac{2\pi kT}{m\omega^2} \right)^{1/2} \left[ 1 - \frac{3\mu^2}{4} + O(\mu^4) \right] \qquad (2.2)$$

The free energy is known to be related to $Z_N$ through $F = -kT \ln Z_N = -NkT \ln Z$ and hence from Eq. (2.2) we obtain to the lowest non-vanishing order in $\mu$

$$F = -NkT \left[ \ln\left( \frac{2\pi kT}{h\omega} \right) - \frac{3\lambda kT}{4m^2\omega^4} \right] \qquad (2.3)$$

The molar specific heat $C_V = -T \dfrac{\partial^2 F}{\partial T^2}$ follows as

$$C_V = R \left[ 1 - \frac{3\lambda kT}{2\left( m\omega^2 \right)^2} + \dots \right] = R \left[ 1 - \frac{3}{2} \mu^2 + \dots \right] \qquad (2.4)$$

On the other hand, if $\mu \gg 1$, we have

$$Z = \int_{-\infty}^{+\infty} \frac{dp}{h} e^{-(p^2/2mkT)} \int_{-\infty}^{\infty} dx\, exp\left[ -\left( \frac{\lambda x^4}{4kT} + \frac{m\omega^2 x^2}{2kT} \right) \right]$$

$$= \left( \frac{2\pi mkT}{h^2} \right)^{1/2} \int_{-\infty}^{\infty} dx\, e^{-\left( \lambda x^4 / 4kT \right)} \left[ 1 - \frac{m\omega^2 x^2}{2kT} + O(\omega^4) \right]$$

$$= \frac{1}{2} \Gamma\left( \frac{1}{4} \right) \left( \frac{2\pi mkT}{h^2} \right)^{1/2} \left( \frac{4kT}{\lambda} \right)^{1/4} \left[ 1 - \frac{\Gamma(3/4)}{\Gamma(1/4)} \frac{1}{\mu} + \dots \right] \qquad (2.5)$$

As before we use the connections between the partition function and the free energy followed by that between the free energy and the specific heat to obtain

$$C_V = \frac{3R}{4} + \frac{R}{4}\frac{\Gamma(3/4)}{\Gamma(1/4)}\frac{1}{\mu} + O\left(\mu^{-2}\right) \qquad (2.6)$$

Our goal will be to obtain Eq. (2.4) and (2.6) starting from the microcanonical ensemble.

### III.  ARBITRARY RESTORING FORCES AND PERTURBATION THEORY IN THE MICROCANONICAL ENSEMBLE

We divide this section into three parts. In part A, we show how the entropy and hence the specific heat of an oscillator whose restoring force is proportional to $x^{2n-1}$ can be calculated in the microcanonical ensemble. In part B, we treat the anharmonic oscillator governed by the Hamiltonian shown in Eq. (1.5) in the limit where the quadratic term dominates. In part C, we sketch how the calculation is done in the case when the quartic term in Eq. (1.5) dominates.

### A) The case of $\phi(x) = Kx^{2n}$

We want to calculate the exact specific heat of $N$ oscillators confined by the potential $\phi(x) = Kx^{2n}$ ($n$ is an integer). The total energy of the system is given by

$$E = \sum_{i=1}^{N}\frac{p_i^2}{2m} + K\sum_{i=1}^{N}x_i^{2n} \qquad (3.1)$$

We use Eq. (1.2) to obtain the number of accessible states available to the system. Clearly for a total energy $E$

$$\Gamma(E,N) = \frac{1}{h^N}\int dx_1 dx_2....dx_N \int dp_1 dp_2...dp_N \delta(E - \sum_{i=1}^{N}[\frac{p_i^2}{2m} + Kx_i^{2n}])$$

$$= \frac{1}{h^N}\int dx_1......dx_N \int dp_1...dp_N \delta(\bar{E} - \sum^{N}\frac{p_i^2}{2m}) \qquad (3.2)$$

In the above $\bar{E} = E - K\sum_{i=1}^{N}x_i^{2n}$ . For a given $\bar{E}$ , the integral over the $p$ - variables gives the surface area of a N-dimensional sphere of radius $\sqrt{2m\bar{E}}$ . This area is $S_N\left(R = \sqrt{2m\bar{E}}\right) = 2\pi^{N/2}(2m\bar{E})^{(N-1)/2}/\Gamma(N/2)$. Since we will always deal with a very large number of particles ( $N \sim 10^{20}$ ) it makes no difference if we replace $N-1$ by $N$ in the above expression for $S_N$ and this is an approximation that we will always use in the remainder of the paper. With this understanding we now write Eq.(3.2) as

$$\Gamma(E,N) = \left(\frac{2m}{h^2}\right)^{N/2} S_N \int dx_1 dx_2 ... dx_N \left(E - K \sum_{i=1}^{N} x_i^{2n}\right)^{N/2} \qquad (3.3)$$

In the above equation we have used $S_N = 2\pi^{N/2}/\Gamma(N/2)$ as the surface area of a unit sphere in N dimensions.

The x-integrations in Eq. (3.3) are best done step by step. We define a sequence of energies by the relation

$$E_j = E - K \sum_{i=j+1}^{N} x_i^{2n} \qquad (3.4)$$

The integer runs from 1 to $N-1$ with the sequence ending with $E_{N-1} = E - Kx_N^{2n}$. We now carry out sequentially the integrations over $x_1, x_2, ... x_N$. This yields

$$\Gamma(E,N) = \left(\frac{2m}{h^2}\right) S_N \int dx_N dx_{N-1} .... dx_2 \int dx_1 \left(E_1 - Kx_1^{2n}\right)^{N/2}$$

$$= \frac{1}{nK^{1/2n}} \left(\frac{2m}{h^2}\right)^{N/2} S_N \beta\left(\frac{N}{2}+1, \frac{1}{2n}\right) \int dx_N ... dx_3 \int dx_2 \left(E_2 - Kx_2^{2n}\right)^{\frac{N}{2}+\frac{1}{2n}} \qquad (3.5a)$$

after the first integration. Carrying out the successive integrations similarly, one has

$$\Gamma(E,N) = S_N \left(\frac{2m}{h^2}\right)^{N/2} \left(\frac{1}{nK^{1/2n}}\right)^N \beta\left(\frac{N}{2}+1, \frac{1}{2n}\right)\beta\left(\frac{N}{2}+1+\frac{1}{2n}, \frac{1}{2n}\right)\beta\left(\frac{N}{2}+1+\frac{2}{2n}, \frac{1}{2n}\right)...$$

$$...... \beta\left(\frac{N}{2}+1+\frac{N-1}{2n}, \frac{1}{2n}\right) E^{\frac{N}{2}+\frac{N}{2n}} \qquad (3.5b)$$

The string of beta functions simplifies to (for $N >> 1$)

$$\frac{\Gamma\left(\frac{N}{2}+1\right)\Gamma\left(\frac{1}{2n}\right)}{\Gamma\left(\frac{N}{2}+1+\frac{1}{2n}\right)}\frac{\Gamma\left(\frac{N}{2}+1+\frac{1}{2n}\right)\Gamma\left(\frac{1}{2n}\right)}{\Gamma\left(\frac{N}{2}+1+\frac{2}{2n}\right)}\frac{\Gamma\left(\frac{N}{2}+1+\frac{2}{2n}\right)\Gamma\left(\frac{1}{2n}\right)}{\Gamma\left(\frac{N}{2}+1+\frac{3}{2n}\right)}....\frac{\Gamma\left(\frac{N}{2}+1+\frac{N-1}{2n}\right)\Gamma\left(\frac{1}{2n}\right)}{\Gamma\left(\frac{N}{2}+1+\frac{N}{2n}\right)}$$

$$=\frac{\left[\Gamma\left(\frac{1}{2n}\right)\right]^N \Gamma\left(\frac{N}{2}+1\right)}{\Gamma\left(\frac{N}{2}\left(1+\frac{1}{n}\right)+1\right)} \tag{3.5c}$$

The phase space volume becomes (in the tremendously large $N$ limit)

$$\Gamma(E,N) = N\left(\frac{2\pi m \Gamma^2(1/2n)}{n^2 h^2 K^{1/n}}\right)^{N/2}\left(\frac{eE}{N}\right)^{N(n+1)/2n} \tag{3.6}$$

The entropy follows as $S = k \ln\Gamma(E,N) = f(N) + kN\left(\frac{1}{2}+\frac{1}{2n}\right)\ln E$ and the extensive nature of entropy is clearly seen when we take a logarithm in both sides of Eq.(3.6). It is seen that the right side contains $O(N)$ terms as dominating order. Note, the function $f(N)$ is that part of $\ln\Gamma(E,N)$ that has no dependence on $E$.

Using Eq. (1.4), we find $E = N\left(\frac{1}{2}+\frac{1}{2n}\right)kT$ and hence a molar specific heat of

$C_V = \frac{R}{2n}(1+n)$. For $n = 1$(simple harmonic oscillator), we get the expected

$C_V = R$ and for the quartic oscillator ($n = 2$) we get $C_V = 3R/4$. The issue of extensivity of entropy will play an important role in our subsequent discussion.

### B) ANHARMONIC OSCILLATOR WITH A WEAK QUARTIC TERM

In this section we turn to the potential $\phi(x) = \frac{1}{2}m\omega^2 x^2 + \frac{\lambda}{4}x^4$ and consider the usual situation where the quartic term is much smaller than the quadratic one. The phase space volume and from it the number of microstates is calculated by evaluating the integral (using the result of the last section)

$$\Gamma(E,N) = \frac{1}{h^N}\int dx_1...dx_N \int dp_1.....dp_N \ \delta\left(E - \sum_{i=1}^{N}(\frac{p_i^2}{2m}+\phi(x_i))\right)$$

$$= \left(\frac{2m}{h^2}\right)^{N/2} S_N \int dx_1 \ldots dx_N \left(E - \sum_{i=1}^{N}\left[\frac{1}{2}m\omega^2 x_i^2 + \frac{\lambda}{4}x_i^4\right]\right)^{N/2} \qquad (3.7)$$

It is convenient to use a set of dimensionless spatial coordinates $y_i$ defined by

$$y_i = \sqrt{\frac{m\omega^2 N}{2E}}\, x_i \qquad (3.8)$$

In terms of these variables Eq. (3.7) becomes

$$\Gamma(E,N) = \left(\frac{2}{h\omega}\right)^N \left(\frac{E}{N}\right)^{N/2} S_N \int dy_1 \ldots dy_N \left(E - \frac{E}{N}\sum_{i=1}^{N}\left[y_i^2 + \frac{\lambda E}{N\left(m\omega^2\right)^2}y_i^4\right]\right)^{N/2}$$

$$= S_N N^{-N/2}\left(\frac{2E}{h\omega}\right)^N \int dy_1 \ldots dy_N \left(1 - \frac{1}{N}\sum_{i=1}^{N}\left[y_i^2 + \bar{\mu}^2 y_i^4\right]\right)^{N/2} \qquad (3.9)$$

In the above equation $\bar{\mu}^2 = \lambda E / \left(m\omega^2\right)^2 N$ which is an intensive quantity. We expand the integrand of Eq. (3.9) in powers of $\bar{\mu}^2 / N$.

To proceed further, we define $y_i = \sqrt{N} z_i$ and obtain after expanding and keeping only the next to leading term (first term in perturbation theory)

$$\Gamma(E,N) = S_N\left(\frac{2E}{h\omega}\right)^N \int_0^1 dz_1 \ldots \int_0^1 dz_N \left(1 - \sum_{i=1}^{N} z_i^2\right)^{N/2} \times \left[1 - \frac{\bar{\mu}^2 N^2}{2}\sum_{j=1}^{N}\frac{z_j^4}{\left(1 - \sum_{i=1}^{N} z_i^2\right)} + \ldots\right] \qquad (3.10)$$

To do the integrals in the above equation, one thinks of $z_1, z_2, \ldots z_N$ as the components of a N-dimensional vector whose magnitude lies between 0 and 1. The quantity $dz_1 \ldots dz_N$ is the infinitesimal volume element of a sphere in N-dimensions. It has the form $S_N r^{N-1} dr$ where $r^2 = \sum_{i=1}^{N} z_i^2$ i.e., $r$ is the radius of the sphere with $0 \le r \le 1$. The integrals in Eq. (3.10) become integrals over the volume of the unit sphere. The first integral can be written as

$$I_1(N) = \int_0^1 dz_1 .. \int_0^1 dz_N \left(1 - \sum_{i=1}^N z_i^2\right) = S_N \int_0^1 r^{N-1} \left(1 - r^2\right)^{N/2} dr \qquad (3.11a)$$

For the second we define $I_2(N) = \int_0^1 dz_1 .. \int_0^1 dz_N \sum_{j=1}^N \frac{z_j^4}{2} \left(1 - \sum_{i=1}^N z_i^2\right)^{\frac{N}{2}-1}$ and obtain

$$I_2(N) = N S_N \frac{\Gamma(N/2)}{\Gamma((N-1)/2)\Gamma(1/2)} \int_0^1 dr r^{N-1} \int_0^\pi d\theta \sin^{N-2}\theta \cos^4\theta r^4 \left(1 - r^2\right)^{(N-2)/2} \quad (3.11b)$$

The factor of $N$ in Eq. (3.11b) comes from the sum over "j" in Eq. (3.10). The integrals evaluate to

$$I_1(N) = \frac{S_N}{4} \frac{\Gamma(N/2)^2}{\Gamma(N)} \qquad (3.12a)$$

$$I_2(N) = \frac{N}{2} S_N \frac{\Gamma(5/2)}{\Gamma(1/2)} \frac{\left[\Gamma(N/2)\right]^2}{\Gamma(N+2)} \qquad (3.12b)$$

We can now write the number of available states in Eq. (3.10) as

$$\Gamma(E,N) = \frac{S_N^2}{2} N^{N/2} \left(\frac{2E}{h\omega}\right)^N \frac{\Gamma\left(\frac{N}{2}+1\right)\Gamma\left(\frac{N}{2}\right)}{\Gamma(N+1)} \left[1 - \frac{3\bar{\mu}^2}{8} N^3 \frac{\Gamma\left(\frac{N}{2}\right)\Gamma(N+1)}{\Gamma\left(\frac{N}{2}+1\right)\Gamma((N+2))}\right]$$

$$= \frac{S_N^2}{2} \left(\frac{2E}{h\omega}\right)^N \frac{\Gamma^2(N/2)}{\Gamma(N+1)} \left[1 - \frac{3}{4}\bar{\mu}^2 N\right] \qquad (3.13)$$

In arriving at the last line, differences between $N$ and $N+1$ and $N/2$ and $(N/2)+1$ have been ignored. The above expression needs to be exponentiated to yield the extensive property of entropy and at $O(\bar{\mu})$, the necessary exponentiation gives

$$\Gamma(E,N) = \frac{S_N^2}{2} \left(\frac{2E}{h\omega}\right)^N \frac{\Gamma^2(N/2)}{\Gamma(N+1)} \left(1 - \frac{3}{4}\bar{\mu}^2\right)^N \qquad (3.14)$$

The entropy follows as $S = k \ln \Gamma(E,N) = g(N) + kN \ln E - \frac{3}{4} N \bar{\mu}^2 + ..$ correct to the lowest order in $\bar{\mu}^2$.

The temperature can now be calculated from

$$\frac{1}{T} = \frac{\partial S}{\partial E}\bigg|_N = \frac{kN}{E} - \frac{3}{4}\frac{\lambda}{\left(m\omega^2\right)^2} \tag{3.15}$$

To the lowest order in $\lambda$, this yields

$$E = NkT - \frac{3}{4}\frac{\lambda ET}{\left(m\omega^2\right)^2} + O(\lambda^2) = NkT - \frac{3}{4}\frac{\lambda NkT^2}{\left(m\omega^2\right)^2} + O(\lambda^2) \tag{3.16}$$

The specific heat is obtained from $C_V = \frac{\partial E}{\partial T}$ and is in exact agreement with Eq.(2.4). Thus, we see that perturbation theory can indeed be carried out in the microcanonical ensemble. In the next sub-section, we demonstrate that it is possible to do the more difficult case of a weak quadratic term as well.

## C) ANHARMONIC OSCILLATOR WITH A WEAK QUADRATIC TERM

We will now focus on the situation where the quartic term dominates and the quadratic term is a perturbation. We will only sketch the steps as they are very similar to those carried out in the previous sub-section. In the previous case where the quartic term was small, we were perturbing around a sphere in phase space and that kept the integrals particularly simple. Here the zeroth order phase space does not have spherical symmetry and the integrals will need careful handling.

Our starting point is Eq. (3.7) as before. The scaled variable $z$ will now be defined by the relation

$$N^{1/4}z_i = y_i = \left(\frac{\lambda N}{4E}\right)^{1/4}x_i \tag{3.17}$$

leading to (analogues of Eq. s (3.10)-(3.13) )

$$\Gamma(E,N) = S_N\left(\frac{2m}{h^2}\right)^{N/2}\left(\frac{4E}{\lambda}\right)^{N/4}E^{N/2}\int dz_1...dz_N\left[\left(1 - \sum_{i=1}^{N}z_i^4\right)^{N/2} - m\omega^2\sqrt{\frac{N^2}{4\lambda E}}\left(\sum_{i=1}^{N}z_i^2\right)\left(1 - \sum_{i=1}^{N}z_i^4\right)^{\frac{N}{2}-1}\right]$$

$$= S_N \left( \frac{2m}{h^2} \right)^{N/2} \left( \frac{4E}{\lambda} \right)^{N/4} E^{N/2} \left[ J_1 - \frac{m\omega^2}{2\sqrt{\lambda E}} N J_2 \right] \qquad (3.18)$$

The integrals $J_{1,2}$ are given by

$$J_1 = \int\limits_0^1 dz_1 \ldots\ldots dz_N \left( 1 - \sum_{i=1}^N z_i^4 \right) \qquad (3.19a)$$

$$J_2 = \int\limits_0^1 dz_1 \ldots\ldots dz_N \left( \sum_{i=1}^N z_i^2 \right) \left( 1 - \sum_{i=1}^N z_i^4 \right)^{\frac{N}{2}-1} = N \int\limits_0^1 dz_1 \ldots dz_N z_1^2 \left( 1 - \sum_{i=1}^N z_i^4 \right)^{\frac{N}{2}-1} \qquad (3.19b)$$

The integral $J_1$ is integrated step by step exactly as demonstrated in Section IIIA and leads to

$$J_1 = \left( \frac{1}{2} \right)^N \beta\left( \frac{N}{2}+1, \frac{1}{4} \right) \beta\left( \frac{N}{2}+1+\frac{1}{4}, \frac{1}{4} \right) \ldots \beta\left( \frac{N}{2}+1+\frac{N-1}{4}, \frac{1}{4} \right) \qquad (3.20a)$$

$$J_2 = \left( \frac{1}{2} \right)^N \beta\left( \frac{N}{2}, \frac{3}{4} \right) \beta\left( \frac{N}{2}+\frac{3}{4}, \frac{1}{4} \right) \beta\left( \frac{N}{2}+\frac{3}{4}+\frac{1}{4}, \frac{1}{4} \right) \ldots \beta\left( \frac{N}{2}+\frac{3}{4}+\frac{N-2}{4}, \frac{1}{4} \right) \quad (3.20b)$$

The ratio $J_2 / J_1$ works out to be ( for $N \gg 1$ ) $\dfrac{2}{N} \dfrac{\Gamma(3/4)}{\Gamma(1/4)} \sqrt{\dfrac{3N}{4}}$ and Eq.(3.18) becomes

$$\Gamma(E,N) = F(N) E^{3N/4} \left[ 1 - \frac{m\omega^2}{\sqrt{\lambda E}} \sqrt{\frac{3N}{4}} \frac{\Gamma(3/4)}{\Gamma(1/4)} \right] \qquad (3.21)$$

We now follow the usual procedure of using $T^{-1} = \left. \dfrac{\partial S}{\partial E} \right|_N$ to find the energy followed by $C_V = \partial E / \partial T$ at constant N to arrive at Eq. (2.6). This completes the demonstration that in practice the microcanonical ensemble can be used for carrying out perturbative calculations.

## IV.    A CROSSOVER FUNCTION

In this section, we show how the microcanonical ensemble can be used to construct an explicit crossover function for the specific heat of the anharmonic oscillator specified in Eq.(1.5). We will obtain the molar specific heat in the form

$$C(T) = Rf\left(\frac{\lambda kT}{m^2 \omega^4}\right) \qquad (4.1)$$

with the function $f(X)$ having the limiting forms

$$f(X) \to 1, X \to 0$$

$$f(X) \to 3/4, X \to \infty$$

The calculation will involve an approximation but that approximation will enable us to write down a complete crossover function. In the canonical ensemble, to obtain the crossover, one would have to obtain a few terms in the low $X$ approximation and then a few terms in the high $X$ limit. Having a weak-coupling and a strong coupling expansion, one would be able to bridge the two limits by a Pade approximant. The microcanonical approach provides a direct answer and an interesting connection with the dynamics of the system.

We begin with the dynamics of a single particle in the anharmonic potential. This has the form

$$\ddot{x} + \omega^2 x + \frac{\lambda}{m} x^3 = 0 \qquad (4.2)$$

The dynamics alters the frequency of oscillation of the system. The frequency now becomes dependent on the amplitude (and hence the energy) of the system. This correction to the frequency can be worked out using the Lindstedt – Poincare technique (see for e.g., Ref. [19]) which gives the first order frequency shift as

$$\Delta\omega^2 = \frac{3}{4}\frac{\lambda}{m}a^2 = \frac{3}{2}\frac{\lambda E}{N m^2 \omega^2} \qquad (4.3)$$

This is valid for $\lambda E/N m^2 \omega^4 << 1$. Thus, at low values of $\lambda$ ( weak nonlinearity), the frequency of the oscillator is $\Omega^2 = \omega^2 \left(1 + \frac{3\lambda E}{2N m^2 \omega^4} + O(\lambda^2)\right)$. For the quartic oscillator ($\lambda kT >> (m\omega^2)^2$), the frequency $\Omega$ is known to scale as $\lambda^{1/4}$. A crossover function for the oscillator frequency was obtained for this problem [20] by writing the dynamics as that of an effective simple harmonic oscillator with the Hamiltonian $H = \frac{p^2}{2m} + \frac{m\omega^2 x^2}{2} + \frac{\lambda \alpha a^2 x^2}{2}$, where $a$ is the amplitude of motion and $\alpha$ is a number of $O(1)$. This Hamiltonian corresponds to an oscillator of frequency $\Omega$, where

$$\Omega^2 = \omega^2 + \frac{\alpha \lambda a^2}{m} \qquad (4.4)$$

Since the energy $\Sigma$ of the simple harmonic oscillator is related to the amplitude as $\Sigma = m\Omega^2 a^2/2$, and in the small $\lambda$ domain Eq. (4.3) must hold, we can write Eq. (4.4) as

$$\Omega^2 = \omega^2 + \frac{3\lambda \Sigma}{2m^2 \Omega^2} \qquad (4.5)$$

This leads to the frequency energy relation as (only the positive sign in the solution of the quadratic is relevant for $\lambda > 0$, as $\Omega^2$ is always non-negative)

$$\Omega^2 = \frac{\omega^2}{2} + \sqrt{\frac{\omega^4}{4} + \frac{3\lambda \Sigma}{2m^2}} \qquad (4.6)$$

This result is exact to $O(\lambda)$ for small $\lambda$ and is accurate to 8% for very high $\lambda$. In the statistical mechanical problem, the above reduction of the dynamics to an effective quadratic Hamiltonian will be the basis of our calculation. The single particle energy is now $\Sigma = E/N$ and we have a collection of $N$ non-interacting simple harmonic oscillators with each of them having a frequency

$\Omega = \omega \left[ \frac{1}{2} + \sqrt{\frac{1}{4} + \frac{3\lambda E}{2m^2 N \omega^4}} \right]^{1/2}$. The available phase space is given by the expression

$$\Gamma(E) = \int \frac{dp_1 dp_2 \ldots dx_1 dx_2 \ldots}{h^N} \delta\left( E - \sum_{i=1}^{N} \frac{p_i^2}{2m} - \sum_{i=1}^{N} \frac{1}{2} m\Omega^2 x_i^2 \right) \qquad (4.7)$$

Using Eq. (3.6), we write

$$\Gamma(E, N) = \frac{2\pi^N (2mE)^N}{(mh\Omega)^N \Gamma(N)} \qquad (4.8)$$

The approximation $N \gg 1$ has been used wherever applicable. Use of Eq. (4.6) now leads to

$$\Gamma(E) = \frac{1}{\Gamma(N)}\left(\frac{2E}{\hbar\omega}\right)^N \left[\frac{1}{1+\sqrt{1+\frac{6\lambda E}{m^2 N\omega^4}}}\right]^{N/2} \tag{4.9}$$

The entropy follows as $S = k\left(N\ln\left(\frac{2E}{N\hbar\omega}\right) + N - \frac{N}{2}\ln\left[1+\sqrt{1+\frac{6\lambda E}{m^2 N\omega^4}}\right]\right)$.

Using the identity $\frac{\partial S}{\partial E} = T^{-1}$, we arrive at the energy as a function of

temperature (defining a dimensionless nonlinearity parameter: $\mu = \frac{\sqrt{\lambda kT}}{m\omega^2}$)

$$\frac{E}{NkT} = 1 - \frac{3\mu^2 E}{2NkT\left[1+6\frac{\mu^2 E}{NkT}+\sqrt{1+6\frac{\mu^2 E}{NkT}}\right]} \tag{4.10}$$

Some algebra gives $E$ as a function of temperature. We got a cubic equation to solve and used Cardan's method for solving the same (See Ref. [21], for a recent discussion). This gave

$$E = \frac{NkT}{6\mu^2}(\epsilon^2 - 1) \tag{4.11}$$

where,

$$\epsilon = 2\left[\frac{1}{3}+\frac{3\mu^2}{2}\right]^{1/2}\cos\left[\frac{1}{3}\tan^{-1}\left(\frac{\sqrt{4\left(\frac{1}{3}+\frac{3\mu^2}{2}\right)^3-\frac{9}{4}\mu^4}}{3\mu^2/2}\right)\right] \tag{4.12}$$

This gives us a closed form estimation of energy as a function of temperature. To keep things cleaner, we define

$$\gamma = \left[\frac{1}{3}+\frac{3\mu^2}{2}\right] \tag{4.13}$$

so that

$$\epsilon = 2\gamma^{1/2} \cos\left[\frac{1}{3}\tan^{-1}\left(\frac{\sqrt{4\gamma^3 - \frac{9}{4}\mu^4}}{3\mu^2/2}\right)\right] \qquad (4.14)$$

Now we get the specific heat as $C(T) = \partial E / \partial T$:

$$C(T) = \frac{Nk}{3}\epsilon^2\left[\frac{3}{4\gamma} - \tan\left(\frac{1}{3}\tan^{-1}\left(\sqrt{\frac{16\gamma^3}{9\mu^4} - 1}\right)\right) \times \frac{8\gamma^2(\gamma-1)}{27\mu^6} \times \right.$$

$$\left. \frac{1}{\left(\frac{16\gamma^3}{9\mu^4} - 1\right)^{1/2} + \left(\frac{16\gamma^3}{9\mu^4} - 1\right)^{3/2}}\right] \qquad (4.15)$$

For Avogadro number of particles $Nk = R$. We then recognize that

$$f\left(\frac{\lambda kT}{m^2\omega^4}\right) = f(\mu^2) = \frac{1}{3}\epsilon^2\left[\frac{3}{4\gamma} - \tan\left(\frac{1}{3}\tan^{-1}\left(\sqrt{\frac{16\gamma^3}{9\mu^4} - 1}\right)\right) \times \right.$$

$$\left. \frac{8\gamma^2(\gamma-1)}{27\mu^6} \times \frac{1}{\left(\frac{16\gamma^3}{9\mu^4} - 1\right)^{1/2} + \left(\frac{16\gamma^3}{9\mu^4} - 1\right)^{3/2}}\right] \qquad (4.16)$$

This does satisfy the required limiting values. So, we have the crossover function in a closed form.

## V.    THE INTERACTING GAS

The ideal gas that we have discussed so far is characterized by the absence of any inter-particle interaction. In reality the molecules (considered spherical) of the gas are capable of interacting with each other. If we focus on the two – body interactions alone, there will be a very strong short-range repulsion (almost like an elastic collision) and a weak attraction at large distances. A convenient characterization of the inter-particle potential is by the Lennard-Jones potential between atoms '$i$' and '$j$'

$$\phi(r_{ij}) = -V_0 \left[ \left( \frac{r_0}{r_{ij}} \right)^6 - \left( \frac{r_0}{r_{ij}} \right)^{12} \right] \tag{5.1}$$

The length scale $r_0$ is almost the diameter of the molecule. This implies that around the centre of each molecule there is a sphere of exclusion of radius $r_0$ in which no other molecule can have its centre. Thus, the total accessible volume is not the container volume $V$ but the excluded volume $V - b$, where $b = 4\pi N r_0^3 / 3$. If we use this simple but reasonable picture of the effect caused by the repulsive forces at very short distances (for a more detailed discussion of this one can consult Refs. [22] and [23]), the equation of state becomes $P(V - b) = NkT$. To find the effect of the attractive part of the potential, we need to use the results obtained in Sec. III. We begin by writing the analogue of Eq. (3.3) in three spatial dimensions as

$$\Gamma(E, N, V) = \frac{1}{N!} S_{3N} \left( \frac{2m}{h^2} \right)^{3N/2} \int d^3 r_1 d^3 r_2 \ldots d^3 r_N \left[ E - \sum_{pairs} \phi(r_{ij}) \right]^{3N/2}$$

$$\tag{5.2}$$

Notice, now the particles do not have their respective force centres, unlike the external potentials of the earlier problems, rather they are involved in mutual interactions. Therefore the $1/N!$ factor must be included in the calculation. This is also necessary for the entropy to be extensive. The repulsive part of the force can be handled by an excluded volume effect which changes the effective volume to $V - b$ (a more accurate evaluation was done by Saha and Basu [22] and a recent variation of that can be found in Ref. [23]). We will be concerned only with the attractive force over here. It is a weak force of electromagnetic origin and captured by the first term in the Lennard Jones potential of Eq. (5.1). Being weak it is amenable to perturbation theory.

We use the smallness of the potential to expand it as (first order in the potential)

$$\Gamma(E, N, V) = \frac{1}{N!} S_{3N} \left( \frac{2m}{h^2} \right)^{3N/2} \int d^3 r_1 \ldots d^3 r_N E^{3N/2} \left[ 1 - \frac{3N}{2} \sum_{pairs} \frac{\phi(r_{ij})}{E} \right] \tag{5.3}$$

Each pair makes the same contribution to the integral and since the number of pairs is $N(N-1)/2$ which is as good as $N^2/2$, we can write the above equation as

$$\Gamma(E, N, V) = \frac{1}{N!} S_{3N} \left(\frac{2mE}{h^2}\right)^{3N/2} V^N \left[1 - \frac{3N}{4E} \frac{N^2}{V^2} \int d^3r_1 d^3r_2 \phi(r_{12})\right] \qquad (5.4)$$

The integral in the above equation is evaluated by keeping only the attractive part of the interaction i.e., $\phi(r_{12}) = -V_0 \left(r_0 / r_{12}\right)^6$. These yields $\int d^3r_1 d^3r_2 \phi(r_{12}) = V \int d^3r_{12} \phi(r_{12})$. The integration over $r_{12}$ extends from infinity down to the lower cut off of $r_0$ and hence Eq. (5.4) becomes

$$\Gamma(E, N, V) = \frac{1}{N!} S_{3N} V^N \left(\frac{2mE}{h^2}\right)^{3N/2} \left[1 + \pi V_0 \frac{N^2}{V} \frac{N}{E} r_0^3\right] \qquad (5.5)$$

We now exponentiate the expression in the square bracket to maintain the extensive property of entropy and obtain

$$\Gamma(E, N, V) = \frac{1}{N!} S_{3N} V^N \left(\frac{2mE}{h^2}\right)^{3N/2} \left(1 + \frac{2\pi}{3} \frac{N^2}{V} \frac{V_0}{E} r_0^3\right)^{3N/2} \qquad (5.6)$$

The entropy follows as $S = k \ln \Gamma$ and the pressure from $P = T \frac{\partial S}{\partial V}$ at constant $E$ and $N$. Standard manipulations yield $E = 3NkT / 2$ to the lowest order in $V_0 / E$ and

$$P = \frac{NkT}{V} - \frac{\pi NkT}{E} V_0 \frac{N^2}{V^2} r_0^3 \qquad (5.7)$$

Using $E = 3NkT / 2$ to the lowest order to substitute for $E$ in the second term on the right-hand side above and using the excluded volume effect discussed above in the first term of Eq. (5.7) to change $V$ to $V - b$, we get

$$P = \frac{NkT}{V - b} - \frac{2\pi}{3} V_0 \frac{N^2}{V^2} r_0^3 \qquad (5.8)$$

Defining the constant $a$ as $a = 2\pi V_0 r_0^3 N^2 / 3$, we can write Eq. (5.8) as

$$\left(P + \frac{a}{V^2}\right)(V - b) = NkT \qquad (5.9)$$

Thus, we have arrived at Van der Waals equation! More correctly we have laid the foundation for the virial expansion via the microcanonical ensemble. We justify this statement in the following discussion.

We extend the binomial expansion of the right-hand side of Eq. (5.3) to write

$$\Gamma = \frac{1}{N!} S_{3N} \left(\frac{2mE}{h^2}\right)^{3N/2} \int d^3r_1 \dots d^3r_N \left[1 - \frac{3N}{2}\sum_{pairs}\frac{\phi(r_{ij})}{E} + \frac{9N^2}{8E^2}\left(\sum_{pairs}\phi(r_{ij})\sum_{pairs}\phi(r_{kl}) + ..\right)\right]$$

(5.10)

The product of the summations in the above equation has three distinct kinds of terms (all counting is done in the limit of $N \gg 1$)

i) The two pairs $ij$ and $kl$ are identical. There are $N^2/2$ terms of this variety

ii) The two pairs have one index in common. These terms are of the form $\phi_{ik}\phi_{il}$ and there $N^3$ terms of this form.

iii) The two pairs have no common index. There are $N^4/4$ terms of this variety.

It is only the third kind of term that needs to be kept in the large-$N$ limit. It follows that at this order we have

$$\Gamma = \frac{1}{N!} S_{3N} \left(\frac{2m}{h^2}\right)^{3N/2} E^{3N/2} V^N \left[1 - \frac{3N}{2E}\frac{N^2}{2V}I + \left(\frac{3N}{2E}\right)^2\frac{N^4}{8V^2}I^2\right]$$

(5.11)

In the above $I$ is the integral $\int d^3r_{12}\phi(r_{12})$. It is easy to check that the above equation corresponds to the first three terms of

$$\Gamma = \frac{1}{N!} S_{3N} \left(\frac{2m}{h^2}\right)^{3N/2} \left(1 - \frac{3N^2}{4EV}I\right)^N$$

(5.12)

and hence corresponds to Van der Waals equation again.

The deviation from Van der Waals equation comes at the next order in the density $n = N/V$. The deviation comes from the contribution of integrals of the product $\phi(r_{12})\phi(r_{13})\phi(r_{34})$ which is exactly the same as what happens in the corresponding calculation in the canonical ensemble. The integrand can be pictured as three linked bonds 21, 13, 34 and hence belongs to the class of linked cluster expansion. What we have demonstrated is that the microcanonical ensemble generates the linked cluster expansion in its own way and gives the Van der Waals equation till the second order and then deviations set in.

### VI.   THE QUANTUM PARTICLE IN A BOX

In this section we return to the issue of whether the perturbation theory results are the same at every order if the perturbation theory is carried out in a microcanonical and canonical ensemble. The accepted result, as already mentioned, is that the two ensembles give identical answers if the entropy as a function of energy is a concave function. Here, we consider a quantum particle in a one-dimensional box of width $L$, so that the energy eigen-values are

$$E_n = \frac{n^2\pi^2\hbar^2}{2mL^2} \tag{6.1}$$

Our goal is to calculate the specific heat of a non-interacting gas of $N$ such particles at high temperatures. As is well known, at very high temperatures (i.e., temperatures $T$ at which $kT >> \pi^2\hbar^2/2ml^2$), the specific heat reaches the value of $Nk/2$. Our interest is in the first correction to this classical result. We first carry out the calculation in the canonical ensemble.

The canonical partition function is given by

$$Z = \sum_{n=1}^{\infty} e^{-\frac{n^2\pi^2\hbar^2}{2mL^2kT}} = \sum_{n=1}^{\infty} e^{-\frac{n^2\alpha^2}{kT}} \tag{6.2}$$

where $\alpha^2 = \pi^2\hbar^2/2mL^2$. To evaluate the sum to two term accuracy, we use the Euler–Mclaurin expansion to two term accuracy as

$$\sum_{n=1}^{\infty} e^{-\frac{n^2\alpha^2}{kT}} = \int_0^{\infty} dn\, e^{-n^2\alpha^2/kT} - \int_0^1 dn\, e^{-n^2\alpha^2/kT} + \frac{1}{2}e^{-\alpha^2/kT} + \cdots \tag{6.3}$$

Working to two term accuracy, we find the partition function as

$$Z = \frac{\sqrt{\pi kT}}{2\alpha} - \frac{1}{2} + O(\alpha^2/kT) \tag{6.4}$$

Using $F = -NkT\ln Z$ where $F$ is the thermodynamic free energy, we obtain the specific from the standard thermodynamic relation as

$$C = \frac{Nk}{2}\left(1 + \frac{\alpha}{2\sqrt{\pi kT}} + \ldots\right) = \frac{Nk}{2}\left[1 + \frac{\sqrt{\pi}\hbar}{2\sqrt{2mL^2kT}} + \ldots\right] \tag{6.5}$$

The first correction obtained above is actually positive! The subsequent terms are negative and lead to a vanishing specific heat as $T \to 0$.

The microcanonical ensemble result, as can be easily checked, is more 'conventional' and gives a negative correction. The microcanonical calculation proceeds by finding the volume of the allowed phase space. If there are $N$ particles with a total energy $E$, then the volume of the phase space is the number of ways $N$ integers ($n_1, n_2, \ldots n_N$) can be constructed so that the sum of squares satisfies

$$\sum_{i=1}^{N} n_i^2 = \frac{2mL^2E}{\pi^2 \hbar^2} \qquad (6.6)$$

For extremely large $E$ and $N$ ( thermodynamic limit ) the integers can be taken to be continuously distributed and the required phase space is $2^{-N}$ th fraction of the volume $V_N$ of the $N$-dimensional sphere of radius $\sqrt{2mL^2E/\pi^2\hbar^2}$. Consequently,

$$\Gamma(E,N) = 2^{-N}V_N = I(N)(2mE/\hbar^2 L^2)^{N/2} \qquad (6.7)$$

The factor $I(N)$ is given by $2^{-N+1}\pi^{N/2}/N\Gamma(N/2)$. The entropy follows as $S = k \ln \Gamma(E,N)$ and $\partial S / \partial E = T^{-1}$ leads to $E = NkT/2$ giving the specific heat as $Nk/2$ - the first term of Eq. (6.5).

We now discuss the correction. The largest source of correction is the elimination of all terms in which at least one $n_i$ happens to be zero. Each $n_i = 0$ defines a plane and hence the correction term for a given $i$ is the fraction of a volume of a $N-1$ dimensional sphere. Since $i$ runs from $1$ to $N$, there are $N$ such volumes and the available phase space is

$$\Gamma(E,N) = V_N - NV_{N-1} = I(N)\left(\frac{2mE}{\hbar^2 L^2}\right)^{N/2}\left(1 - N\frac{I_{N-1}}{I_N}\frac{\hbar L}{\sqrt{2mE}}\right) \qquad (6.8)$$

We are interested in the very large $N$ limit, where $N-1 \cong N$, but $\Gamma(N/2)/\Gamma((N-1)/2) \cong \sqrt{N/2}$ . This casts Eq. (6.8) in the form

$$\Gamma(E,N) = I(N)\left(\frac{2mE}{\hbar^2 L^2}\right)^{N/2}\left(1 - N\hbar L\left(\frac{N}{2mE}\right)^{1/2}\right) \qquad (6.9)$$

Working to the lowest order in the correction to the first term in the bracket,

$$S = k\ln\Gamma(E,N) = \frac{kN}{2}\ln\frac{2mE}{\hbar^2 L^2} - kN\hbar L\left(\frac{N}{2\pi mE}\right)^{1/2} + f(N) \qquad (6.10)$$

A derivative with respect to the energy $E$ takes us to

$$\frac{E}{kT} = \frac{N}{2} + \frac{\hbar L}{2\sqrt{2\pi m}}\frac{N^{3/2}}{E^{1/2}} \qquad (6.11)$$

Since the second term on the right-hand side is much smaller than the first, we can write Eq. (6.11) as

$$E = \frac{NkT}{2} + \frac{N\hbar L}{2\sqrt{\pi m}}\sqrt{kT} \qquad (6.12)$$

Consequently, the specific heat to two term accuracy is given by

$$C = \frac{Nk}{2} - \frac{Nk\hbar L}{4\sqrt{\pi mkT}} \qquad (6.13)$$

The corrections to the leading term in the two ensembles (Eqs (6.5) and (6.13)) differ not only in magnitude but also in sign.

We have an unexpected result in this difference between the canonical and microcanonical approach to calculating a physical quantity in a quantum system. If one does the same calculation for a simple harmonic oscillator, there is complete agreement. The difference seems to be in the existence of a natural scale in the quantum harmonic oscillator – a scale emerging from the combination of $\hbar$, the mass and the "coupling constant" of the harmonic force (the frequency for a simple harmonic oscillator and $\lambda$ for a quartic oscillator characterized by the potential energy $\lambda x^4 / 4$). The absence of a natural scale in the particle in a box (the scale $L$ is associated with the boundary condition) appears to be the reason for this breakdown of equivalence in perturbation theory results in the two ensembles.

It is interesting to note that this drastic difference in perturbation theoretic results for this quantum problem shows the dangers associated with the definition of the thermodynamic limit. Thermodynamic limit has tacitly been assumed to be the number of particles $N \to \infty$ and the volume $V \to \infty$ with the ratio $n = N / V$ remaining finite. This is not allowed in the above quantum problem, where $L \to \infty$ is the classical limit. The perturbation theory calculation of the specific heat of this quantum system in the two ensembles raises a question about ensemble equivalence.

## VII.    CONCLUSION

What we have demonstrated above is the fact that the microcanonical ensemble can be directly used for carrying out a perturbation theoretique calculations in statistical mechanics. This is generally not demonstrated in text books presumably because in the calculation of the phase space volume the various calculated quantities come raised to the power N. Observable quantities require taking a logarithm which makes them numbers of the order of N. What was important was that one has to realize that to the order of perturbation theory that one is carrying out, the phase space volume has to be appropriately exponentiated to make the entropy an extensive quantity. This allows one to do perturbation expansions with confidence.

After having carried out the perturbative calculation for the specific heat of the anharmonic oscillator (potential ...x^2 +....x^4) starting with both lambda=0 and omega=0, we have looked at how to bridge the two limits by a crossover function. This has been done by using a bridging relation from the dynamics of the problem and is a technique that can be effectively used in the microcanonical ensemble. We have also considered the issue of the real gas and shown how Van der Waals equation of state can be obtained in the microcanonical ensemble and also indicated how the linked cluster expansion can be formulated in this case. Finally, we have considered a one-dimensional gas of free quantum particles in the dilute gas limit (i.e., the statistics is not considered) and shown that the perturbation theory for the specific heat in the microcanonical and canonical ensembles can differ significantly.  This is somewhat of a surprise.

**Acknowledgments:**

The authors would like to thank Amit Dutta and Sagar Chakraborty for interesting discussions.

<u>REFERENCES</u>